\documentclass[useAMS,usenatbib]{mn2e_mod}
\usepackage{amsmath,amssymb,graphicx,multirow,xspace,color,hyperref,gensymb}
\usepackage{auto-pst-pdf}

\newcommand{\RUNDMC}{{\tt RUN DMC}\xspace}
\newcommand{\MERCURY}{{\tt MERCURY}\xspace}

\newcommand{\B}{\ensuremath{\textup{``b''}}\xspace}
\newcommand{\C}{\ensuremath{\textup{``c''}}\xspace}
\newcommand{\D}{\ensuremath{\textup{``d''}}\xspace}
\newcommand{\E}{\ensuremath{\textup{``e''}}\xspace}
\newcommand{\F}{\ensuremath{\textup{``f''}}\xspace}

\newcommand{\Rstar}{\ensuremath{R_{\star}}\xspace} 
\newcommand{\Mstar}{\ensuremath{M_{\star}}\xspace} 
\newcommand{\Mearth}{\ensuremath{M_{\oplus}}\xspace} 
\newcommand{\Mjupiter}{\ensuremath{M_J}\xspace} 
\newcommand{\jitter}{\ensuremath{\sigma_{jit}}\xspace}
\newcommand{\Teff}{\ensuremath{T_{eff}}\xspace}
\newcommand{\mutInc}{\ensuremath{i_{mut}}\xspace}

\newcommand{\gcm}{\ensuremath{\textup{g\,cm}^{-3}}\xspace}
\newcommand{\ms}{\ensuremath{\textup{m\,s}^{-1}}\xspace}

\title[55 Cancri]{The 55 Cancri Planetary System: Fully Self-Consistent N-body Constraints and a Dynamical Analysis}

\author[B.~Nelson~et~al.]{
Benjamin~E.~Nelson$^{1,2,3}$\footnotemark[1],
\,Eric~B.~Ford$^{1,2,3}$,
\,Jason~T.~Wright$^{1,2}$,
\,Debra~A.~Fischer$^{4}$,\newauthor
\,Kaspar~von~Braun$^{5}$,
\,Andrew~W.~Howard$^{6}$,
\,Matthew~J.~Payne$^{3,7}$,
\,Saleh Dindar$^{8}$\\
\\
$^{1}$Center for Exoplanets and Habitable Worlds, The Pennsylvania State University, 525 Davey Laboratory, University Park, PA, 16802, USA\\
$^{2}$Department of Astronomy and Astrophysics, The Pennsylvania State University, 525 Davey Laboratory, University Park, PA 16802, USA\\
$^{3}$Department of Astronomy, University of Florida, 211 Bryant Space Science Center, Gainesville, FL 32611, USA\\
$^{4}$Department of Astronomy, Yale University, New Haven, CT 06520 USA\\
$^{5}$Max-Planck Institute for Astronomy (MPIA), K\"{o}nigstuhl 17, 69117 Heidelberg, Germany\\
$^{6}$Institute for Astronomy, University of Hawaii, 2680 Woodlawn Drive, Honolulu, HI 96822, USA\\
$^{7}$Harvard-Smithsonian Center for Astrophysics, 60 Garden Street, Cambridge, MA 02138, USA\\
$^{8}$Department of Computer \& Information Science \& Engineering, University of Florida, CSE Building, Gainesville, FL, 32611, USA\\
}

\begin{document}

\date{Accepted ... Received ...; in original form ...}
\pagerange{\pageref{firstpage}--\pageref{lastpage}} \pubyear{2014}
\maketitle
\label{firstpage}

\begin{abstract}
We present an updated study of the planets known to orbit 55 Cancri A using 1,418 high-precision radial velocity observations from four observatories (Lick, Keck, Hobby-Eberly Telescope, Harlan J. Smith Telescope) and transit time/durations for the inner-most planet, 55 Cancri \E \citep{Winn11}.
We provide the first posterior sample for the masses and orbital parameters based on self-consistent n-body orbital solutions for the 55 Cancri planets, all of which are dynamically stable (for at least $10^8$ years).
We apply a GPU version of Radial velocity Using N-body Differential evolution Markov Chain Monte Carlo (\RUNDMC; \citet{Benelson14}) to perform a Bayesian analysis of the radial velocity and transit observations.

Each of the planets in this remarkable system has unique characteristics.
Our investigation of high-cadence radial velocities and priors based on space-based photometry yields an updated mass estimate for planet \E ($8.09\pm0.26$ \Mearth), which affects its density ($5.51\pm^{1.32}_{1.00}$ \gcm) and inferred bulk composition.
Dynamical stability dictates that the orbital plane of planet \E must be aligned to within 60$\degree$ of the orbital plane of the outer planets (which we assume to be coplanar).
The mutual interactions between the planets \B and \C may develop an apsidal lock about 180$\degree$.
We find 36-45\% of all our model systems librate about the anti-aligned configuration with an amplitude of $51\degree\pm^{6\degree}_{10\degree}$.
Other cases showed short-term perturbations in the libration of $\varpi_b-\varpi_c$, circulation, and nodding, but we find the planets are not in a 3:1 mean-motion resonance.
A revised orbital period and eccentricity for planet \D pushes it further toward the closest known Jupiter analog in the exoplanet population.
\end{abstract}


\footnotetext[1]{e-mail: \rm{\url{benelson@psu.edu}.}}

\section{Introduction}
With roots in early Doppler surveys, 55 Cancri is a wide visual binary system harboring five known planets with a large range of orbital periods ($\sim$0.7 days to $\sim$14 years) and masses ($\sim$8\Mearth to $\sim$4\Mjupiter).
The known planets orbit 55 Cancri A, a K0 type dwarf \citep{vonBraun11}, which is also orbited by an M-dwarf at a projected separation of $\sim$1065AU \citep{Mugrauer06}.
The system has an extensive radial velocity (RV) history using several ground-based facilities (e.g. Lick, Keck, Hobby-Eberly Telescope, Harlan J. Smith Telescope) as well as space-based observatories to constrain properties of the inner-most planet.
The first eight years of RV measurements from the Lick Observatory showed a strong periodic signal with a period of 14.6 days indicative of a $\sim$0.8 \Mjupiter mass planet named 55 Cancri b \citep{Butler97}.
The RV time series showed other trends due to the presence of two additional massive bodies, 55 Cancri \C and \D with orbital periods 44.3 and 5360 days respectively, which were eventually uncovered via additional RV measurements \citep{Marcy02}.
They found the difference between a Keplerian and self-consistent Newtonian fit for a three-planet model is measurable on the observing timescale, demonstrating that the Keplerian model is insufficient for fitting future observations.
RV observations with the Hobby-Eberly Telescope (HET) uncovered a 2.8 day signal of a $\sim$17\Mearth planet, initially labeled as \E \citep{McArthur04}.
However in 2010, a re-analysis of these data showed the previously published orbital period of planet \E was an alias and the more likely period was 0.7365 days \citep{DawsonFabrycky10}.
The revised ephemeris pushed the mass of \E toward the super-Earth regime and raised its transit probability to $\sim$25\%.
Space-based searches with MOST \citep{Winn11} and warm Spitzer \citep{Demory11} showed that planet \E did indeed transit.
More photometric measurements have refined the planet-to-star radius ratio estimate \citep{Demory12, Dragomir13, Gillon12}.
A fifth planet \F was found with an orbital period of 260 days \citep{Fischer08}.
Additional RVs provided a new mass estimate of planet \E and an updated model for the planetary system assuming no planet interactions \citep{Endl12}.

The complexity of the 55 Cancri system provides valuable information for theoretical investigations, improving our understanding of planetary migration, orbital evolution, and composition.
The near 3:1 period commensurability of planets \B and \C is thought to provide evidence for the planets migrating to their present locations, rather than forming in-situ \citep{Kley04}.
The orbital evolution of this planet pair has been a longstanding problem, especially in the presence of additional planets.
It was thought that the three resonant arguments for planet's \B and \C librated, suggesting they orbited in a mean-motion resonance \citep{Ji03,Zhou04,Barnes07}.
With additional RV information, the resonant arguments were found to be circulating \citep{Fischer08} for the single self-consistent dynamical fit, suggesting they were not in resonance.
Injecting a hypothetical sixth planet based on the aforementioned model did not cause libration of these angles \citep{Raymond08}.
The binary companion may play a significant role in the long-term orbital evolution of the system.
Its gravitational perturbations paired with the rapid planet-planet interactions causes the orbits of the outer-four known planets to precess like a rigid body, so they are likely misaligned with the stellar spin axis of 55 Cancri A \citep{Kaib11}.
Improved stellar parameters \citep{vonBraun11} and aforementioned photometric radius estimates for planet \E inspired interior composition models that allow us to learn about a planet that is unlike anything in our own Solar System.
One noteworthy analysis suggests \E is a solid, carbon-rich planet \citep{Madhusudhan12}, a chemical composition much different than that of Earth. 
However, thorough chemical composition models can only be done with extremely precise measurements of the star's radius, C/O ratio, planetary mass, etc. \citep{Teske13}.
The combination of tidal forces and secular interactions with the outer planets are predicted to drive planet \E into a state where its eccentricity oscillates with amplitudes ranging from $10^{-4}$ to 0.17, depending on the orbital parameters and tidal dissipation efficiency \citep{Bolmont13}.
The system as a whole has inspired tests of classical secular theory, which provides insight to the dynamical histories \citep{VanLaerhoven12}.

Characterizing the key physical quantities of this landmark exoplanet system requires a combination of these high-precision measurements (both photometric and spectroscopic) and robust physical and statistical modeling.
Due to limitations in computational power, aforementioned analyses of the RV data assumed each planet travels on an independent Keplerian orbit.
Motivated from the findings of \citet{Marcy02}, \citet{Fischer08} did report a best fit Newtonian model using the period alias for \E but not a detailed analysis of parameter uncertainties with a fully self-consistent dynamical model.
A generalized n-body Markov chain Monte Carlo algorithm applied to RV observations was developed by \citet{Benelson14} to explore complex $\chi^2$ surfaces in high-dimensional parameter spaces.
The 55 Cancri system is a challenging problem due to the length of the RV observing baseline, the sheer number of model parameters, the observed mutual interactions amongst planets, and the short inner orbital period.

In this paper, we present new RVs from the Keck High Resolution Echelle Spectrometer (HIRES), revised constraints on the orbital parameters of the 55 Cancri planets based on a self-consistent dynamical model, and an analysis of the orbital evolution of the system.
In \S\ref{sec:obs}, we describe the Doppler observations used, including a new set made with the Keck HIRES.
In \S\ref{sec:methods}, we briefly describe the \RUNDMC algorithm, the parameter space, and our methods for modeling the observations.
In \S\ref{sec:results}, we present results for each of the 55 Cancri planets.
We conclude with a discussion of the key results and the applications of our posterior samples in \S\ref{sec:discussion}.

\section{Observations}
\label{sec:obs}


\subsection{Lick, HET, and HJST data}
\label{sec:oldRVs}
Our investigation of 55 Cancri began as a performance test of \RUNDMC using the published 70 Keck and 250 Lick RVs from \citet{Fischer08} without considering the inner-most planet \citet{Benelson14}.
We found \RUNDMC could successfully navigate such a high dimensional parameter space ($\sim$5 $\times$ number of planets), so it appeared plausible to perform a much more thorough analysis with all five planets, all published RVs, a dynamical model, and a detailed model of the systematics.
Next, we included the remaining public RVs, specifically from the Hobby-Eberly Telescope (HET) \citep{McArthur04}.
\cite{Endl12} eventually announced new McDonald Observatory RVs, including a new Doppler reduction of the \citet{McArthur04} RVs totaling 131 measurements, and a new set of 212 RVs from the Harlan J. Smith Telescope (HJST).
After the RV program on the Lick observatory's Hamilton spectrograph effectively ended (due to the heater of the iodine cell malfunctioning), \citet{Fischer14} released 582 unbinned RV measurements of 55 Cancri.
These data include a new reduction of the older RVs as well as new observations.
Our final analysis considers RVs from \citet{Endl12} and \citet{Fischer14} RVs, as well as new, unbinned HIRES RVs, to be discussed in the subsequent subsection.
We will describe the treatment of all these observations in \S \ref{sec:modelOfObs}.

\subsection{Keck Data}
\label{sec:newRVs}
Our analysis includes 493 unbinned velocities from Keck HIRES.
These RVs consist of a combination of post-\citet{Fischer08} measurements and individual RVs that were previously reported as one binned measurement.
We present Keck observations in Table \ref{tbl-1}.

We measured relative RVs of 55 Cancri A with the HIRES echelle spectrometer \citep{Vogt94} on the 10-m Keck I telescope using standard procedures.
Most observations were made with the B5 decker (3.5 $\times$ 0.86 arcseconds).
Light from the telescope passed through a glass cell of molecular iodine cell heated to 50\degree C.
The dense set of molecular absorption lines imprinted on the stellar spectra in 5000--6200 $\textup{\AA}$ provide a robust wavelength scale against which Doppler shifts are measured, as well as strong constraints on the instrumental profile at the time of each observation \citep{Marcy92,Valenti95}.
We also obtained five iodine-free ÒtemplateÓ spectra using the B1 decker (3.5 $\times$ 0.57 arcseconds).
These spectra were de-convolved using the instrumental profile measured from spectra of rapidly rotating B stars observed immediately before and after.
We measured high-precision relative RVs using a forward model where the de-convolved stellar spectrum is Doppler shifted, multiplied by the normalized high-resolution iodine transmission spectrum, convolved with an instrumental profile, and matched to the observed spectra using a Levenberg-Marquardt algorithm that minimizes the $\chi^2$ statistic \citep{Butler96}.
In this algorithm, the RV is varied (along with nuisance parameters describing the wavelength scale and instrumental profile) until the $\chi^2$ minimum is reached.
Each RV uncertainty is the standard error on the mean RV of $\sim$700 spectral chunks that are separately Doppler analyzed.
These uncertainty estimates do not account for potential systematic Doppler shifts from instrumental or stellar effects.

\begin{table*}
\caption{\small New, unbinned HIRES velocities for 55 Cancri. \label{tbl-1}}
\begin{tabular}{cccc}
\hline
BJD-2440000. [days] & Radial Velocity [\ms] & Uncertainty [\ms] & Offset Index \\
\hline
\hline
12219.138044 & -147.87 & 1.10 & 0 \\
12236.014734 & -65.34 & 1.21 & 0 \\
12243.050683 & -58.78 & 1.17 & 0 \\
12307.849907 & -111.75 & 1.33 & 0 \\
12333.997674 & -142.98 & 1.49 & 0 \\

\hline
\hline
\end{tabular}

\medskip
Table \ref{tbl-1} is presented in its entirety as Supporting Information with the online version of the article. This stub table is shown for guidance regarding its form and content.
\end{table*}


\section{Methods}
\label{sec:methods}

The Radial velocity Using N-body Differential evolution Markov Chain Monte Carlo code (\RUNDMC; \citet{Benelson14}) was developed to characterize masses and orbits of complex planetary systems with many model parameters.
This algorithm specializes in analyzing RV time series and extracting the orbital parameters assuming an n-body model.
The ``differential evolution'' aspect \citep{terBraak06} helps accelerate the burn-in phase and the mixing of the parameters, especially when covariant structure is present amongst model parameters.
The 55 Cancri system is also ideal for testing the robustness of \RUNDMC for a few reasons: 1) there are many RV observations of 55 Cancri spanning a baseline of $\sim$23 years; 2) this five-planet system requires a minimum of 25 model parameters, excluding on-sky inclination, RV offsets, and jitters; and 3) the dynamical interactions between \B and \C are not negligible on the observing timescale (i.e. not Keplerian).

We developed a version of \RUNDMC that utilizes Swarm-NG to perform n-body integrations using nVidia graphics cards and the CUDA programming environment \citep{Dindar13}.
This allowed us to more effectively parallelize the evolution of many Markov chains on the microprocessors.
On average, the GPU provided a 4-5x speed up over our CPU version.
However, the burn-in phase for some individual test runs exceeded the maximum walltime of 500 hours allowed at the University of Florida High Performance Computing Center (UF HPC).
Our solution was to simply restart such runs from where they stopped until the target distribution was reached.
Still, the generational lag of the autocorrelation was lengthy ($\sim$few thousand) and took on the order of a couple weeks on a GPU to obtain a set of 10,000 effectively independent samples \citep{Benelson14}.
With these posterior samples, we explore the orbital evolution and long-term stability of the system using the hybrid integrator of \MERCURY \citep{Chambers99}. 
This is the first application of our GPU-based \RUNDMC \citep{Dindar13}.

There are a number of input parameters for \RUNDMC.
We considered the lessons from \citet{Benelson14} in order to approach this problem in the most efficient manner.
Extrapolating those results to a 5-planet system with non-negligible self-interactions, we set $n_{chains}=256$, $\sigma_\gamma=0.05$, and MassScaleFactor=1.0.
To accommodate the inner-most planet's 0.7365 day orbital period, we set our integration timestep to $\sim$5 minutes \citep{Kokubo98b}.
We do not consider the wide binary companion, since its perturbations are only significant over very large timescales.

We generated a Keplerian set of initial conditions using a standard random-walk proposal, Metropolis-Hastings MCMC \citep{Ford06}. 
We fed these states into \RUNDMC, which ran for more than 100,000 generations (equivalent to over 25,600,000 model evaluations).
We utilize the resources of the UF HPC for both our CPU and GPU-based computations.
As mentioned in \S \ref{sec:oldRVs}, we began this analysis only considering the \citet{Fischer08} RVs from Lick and Keck.
As we introduced new datasets, we were required to add new parameters describing instrumental properties (i.e. RV offsets and jitters).
We used posterior samples from the previous pilot runs as the initial conditions for parameters shared between both models and synthetically generated initial conditions for every new parameter.
In other words, this study began in a lower dimensional parameter space ($\sim$30) that was eased into a higher dimensional parameter space ($\sim$40) including additional RV data and a more detailed instrumental model (to be discussed in \S \ref{sec:modelOfObs}).


\subsection{Model Parameters}
\label{sec:modelParameters}
We employ \RUNDMC to constrain the Keplerian orbital elements of the 55 Cancri system and the instrumental parameters.
We characterize the system model with the star mass (\Mstar) and radius (\Rstar), plus each planet's mass ($m$), semi-major axis ($a$), eccentricity ($e$),  inclination ($i$), the longitude of periastron ($\omega$), the longitude of ascending node ($\Omega$), and mean anomaly ($M$) at our chosen epoch (first Lick observation) for each planet, plus the RV zero-point offsets ($C$) and jitters (\jitter) for each observatory.
We report the orbital periods ($P$) based on Kepler's Third Law and each body's $m$ and $a$ based in a Jacobi coordinate system.

The 55 Cancri system is well approximated by a coplanar system, i.e. $\Omega=0$ and $i$ is the same for all planets.
Typically, radial velocities do not place strong constraints on $i$ and $\Omega$ unless the self-interactions amongst the planets are very strong (e.g. GJ 876, \citep{Rivera10}; HD82943, \citep{Tan13}).
The near 3:1 MMR in 55 Cancri is significant enough to require an n-body model but we will show it does not provide a strong constraint on orbital inclination of the planets.
However, photometric observations of \E's transit from MOST place a tight constraint on the inclination of this planet.
Thus, we constrain the central transit time, transit duration, and ingress duration based on measurements reported in \citet{Winn11}.
Initially, we assume all orbits are coplanar.
In our final analysis, we allow \E to have a different $i$ and $\Omega$ than the outer four planets, but assume the outer planets are coplanar with each other.

Because the estimates of the orbital parameters of \E now depend on the photometry and thus the radius of the star \Rstar, we adopt the interferometric measurement of \Rstar reported by \citet{vonBraun11}.
Since we are performing n-body integrations to compute the induced RV signal, our initial conditions must specify the mass of each body.
The uncertainty in stellar radius propagates to the stellar mass and further to planet masses.
We consider how \Mstar changes with \Rstar at a constant effective temperature, \Teff, from spectroscopy.
Starting with the mass-luminosity relation $L_\star \sim \Mstar^y \sim \Teff^4 \Rstar^2$, we can derive
\begin{equation}
\left.\frac{\partial \Mstar}{\partial \Rstar}\right|^{}_{\Teff}\sim\frac{2}{y}\frac{\Mstar}{\Rstar}.
\end{equation}
Using theoretical stellar models, we solved for $y$ by adopting the observed \Teff=5196K and $L_\star=0.582 L_\odot$  from \citet{vonBraun11}.
We generated $Y^2$ isochrones \citep{Demarque04} and found the best $y$ value for multiple ages: 4.125 (8Gyr), 4.602 (9Gyr), 5.15 (10Gyr), 5.491 (11Gyr), and 5.817 (12Gyr).
For this analysis, we adopted $y=5.15$ since it corresponds with the age estimate for 55 Cancri \citep{vonBraun11}.
Despite such a relatively steep power law, subsequent test runs showed that our conclusions were not sensitive to whether we allowed for uncertainty in \Mstar described above or assumed a fixed \Mstar value.


\subsection{Model of Observations}
\label{sec:modelOfObs}
Based on preliminary tests, we generalized \RUNDMC to allow for three complications which were not considered in previous analyses.

First, we divided the \citet{Fischer08} Lick dataset into three subsets based on which CCD Dewar was used for each observation.
We found that different Dewars used on the Lick Hamilton spectrograph do not give consistent RV zero-point offsets (Wright, private communication), consistent with results from \citet{Fischer14}.
Therefore, we adopt a different velocity offset depending on which Dewar was being used at the time.
With the \citet{Fischer14} velocities, the offsets were determined by the Dewar codes: 6, 8, 39, 18, and 24.
Dewar code ``6'' has 5 observations ranging from JD-2440000=7578.7300 to 8375.6692 ($C_{1,Lick}$).
Dewar code ``8'' has 12 observations ranging from JD-2440000=8646.0011 to 9469.6478 ($C_{2,Lick}$).
These early Doppler era observations are expected to have relatively high jitter values.
Dewar code ``39'' (Dewar 13 in actuality) has 96 observations ranging from JD-2440000=9676.0632 to 11298.722 ($C_{3,Lick}$).
Dewar code ``18'' (Dewar 6 in actuality) has 91 observations ranging from JD-2440000=11153.033 to 12409.739 ($C_{4,Lick}$).
Dewar code ``24'' (Dewar 8 in actuality) has 378 observations ranging from JD-2440000=12267.957 to 15603.809 ($C_{5,Lick}$) and overlaps with the time series from Dewar code ``18''.
Particle events may have increased the jitter for the latter (Wright, private communication).

Similarly, we split the Keck dataset based on whether observations were taken before or after the CCD upgrade and new Doppler reduction process in 2004.
These two subsets received separate velocity zero-point offset parameters.
Specifically, the pre-CCD upgrade era has 24 observations ranging from BJD-2440000=12219.138044 to 13077.041736 ($C_{1,Keck}$).
The post-CCD upgrade era has 469 observations ranging from BJD-2440000=13339.043299 to 15728.743727 ($C_{2,Keck}$).
We also consider HJST and reanalyzed HET observations provided by \citet{Endl12}.
Each of these datasets has its own offset ($C_{HET}$ and $C_{HJST}$) relative to the large RV zero-points reported by \cite{Endl12}.
In total, we model nine RV offsets.

\begin{table}
\centering
\caption{Definitions for Case 1 and Case 2. In choosing a set of models for subsequent analyses, we recommend using Case 2. \label{tbl-2} }
\begin{tabular}{cc}
\hline
Term & Definition \\
\hline
\hline
Case 1 & all observation errors as uncorrelated \\
\hline
\multirow{2}{*}{Case 2} & errors of ``back-to-back'' observations \\
 & are perfectly correlated \\
\hline
\hline
\end{tabular}

\medskip
Further details are addressed in \S \ref{sec:modelOfObs}.
\end{table}

Second, we include multiple jitter parameters, \jitter, one for each observatory.
Jitter models scatter in observations beyond what is expected from the formal measurement uncertainties.
Jitter may be due to astrophysical noise (e.g. p-modes or chromospheric activity on the star) and/or unmodeled instrumental effects.
We performed preliminary analyses using various combinations of our four datasets (e.g. Keck only, Keck+HET, Keck+Lick, Keck+HET+Lick, etc.).
We found that introducing the Lick dataset increased our jitter estimate.
Furthermore, we expect that \jitter varied within the Lick time series itself \citep{Fischer14}.
Therefore, we define three Lick jitter terms for Dewar codes 6 and 8 ($\sigma_{jit, Lick1}$), 39 and 18 ($\sigma_{jit, Lick2}$), and 24 ($\sigma_{jit, Lick3}$).
We also assign a jitter for Keck ($\sigma_{jit, Keck}$), HET ($\sigma_{jit, HET}$), and HJST ($\sigma_{jit, HJST}$), totaling six jitter terms.
For each observation, we substitute the appropriate jitter term for $\sigma_{jit}$ in our likelihood function (Equations 4 and 5, \citet{Benelson14}).
We modified our likelihood function to include both RV observations and light curve parameters measured from the transit light curve observations for planet \E.
Our $\chi^2_{eff}$ from \citet{Benelson14} is adjusted as such:
\begin{eqnarray}
\chi_{eff}^2 &=&\chi^2 + \sum_k \ln\left[\frac{\sigma_{\star, obs}(t_k,j_k)^2+\sigma_{jit}^2}{\sigma_{\star, obs}(t_k,j_k)^2}\right]\nonumber \\
&& +  \sum_{TT, d_t, d_{in}, R_\star}\frac{(X-X_{obs})^2}{\sigma^2_{X}}
\end{eqnarray}
where $TT$, $d_t$, $d_{in}$, and $R_\star$ are the transit time, transit duration, ingress duration, and stellar radius respectively.
From \citet{Winn11}, we use $TT = 2,455,607.05562\pm0.00087\textup{ HJD}$, $d_t = 0.0658\pm0.0013\textup{ days}$, and $d_{in} = 0.00134\pm0.00011\textup{ days}$.
From \citet{vonBraun11}, we use $R_\star = 0.943\pm0.010$ $R_\odot$.

\begin{table*}
\caption{\small Orbital parameter estimates for all the known 55 Cancri planets from self-consistent dynamical fits. \label{tbl-3} }
\begin{tabular}{cccccc}
\hline
Parameter & Planet e & Planet b & Planet c & Planet fÊ& Planet d \\
\hline
\hline
\multirow{2}{*}{$P$ [days]} &  $0.7365478 \pm^{0.0000014}_{0.0000011}$  &  $ 14.65276 \pm^{0.00082}_{0.00089} $  &  $ 44.380 \pm^{0.020}_{0.018} $  &  $ 261.04 \pm 0.37 $  &  $ 4872 \pm^{28}_{24} $  \\
 &  $0.7365478 \pm^{0.0000016}_{0.0000012}$  &  $14.65314 \pm^{0.00090}_{0.00095}$  &  $44.373 \pm^{0.020}_{0.018}$  &  $260.91 \pm 0.36$  &  $4867 \pm^{25}_{26}$  \\ \hline
\multirow{2}{*}{$P_{avg}$ [days]} & $0.73655015 \pm^{0.00000093}_{0.00000212}$  &  $14.651248 \pm^{0.000084}_{0.000088}$  &  $44.4120 \pm^{0.0050}_{0.0052}$  &  $261.04 \pm 0.37$  &  $4872 \pm^{28}_{24}$  \\
& $0.73655012 \pm^{0.00000099}_{0.00000177}$  &  $14.651248 \pm^{0.000087}_{0.000094}$  &  $44.4094 \pm 0.0055$  &  $260.91 \pm0.36$  &  $4867 \pm^{25}_{26}$  \\ \hline
\multirow{2}{*}{$K$ [\ms]} &  $ 6.04 \pm 0.19$  &  $ 71.60 \pm^{0.19}_{0.21} $  &  $ 10.46 \pm^{0.18}_{0.19} $  &  $ 4.75 \pm 0.19$  &  $ 47.03 \pm^{0.40}_{0.41} $  \\
 &  $6.12 \pm 0.20 $  &  $71.47 \pm 0.21$  &  $10.48 \pm 0.21$  &  $4.80 \pm 0.20$  &  $47.30 \pm^{0.44}_{0.41}$  \\ \hline
\multirow{2}{*}{$m$ [\Mjupiter]} &  $ 0.02513 \pm 0.00079 $  &  $ 0.844 \pm^{0.124}_{0.034} $  &  $ 0.1783 \pm^{0.0261}_{0.0078} $  &  $ 0.1475 \pm^{0.0207}_{0.0093} $  &  $ 3.83 \pm^{0.57}_{0.15} $  \\
 &  $0.02547 \pm^{0.00082}_{0.00081}$  &  $0.840 \pm^{0.131}_{0.031}$  &  $0.1784 \pm^{0.0275}_{0.0078}$  &  $0.1479 \pm^{0.0219}_{0.0093}$  &  $3.86 \pm^{0.60}_{0.15}$  \\ \hline
\multirow{2}{*}{$a$ [AU]} &  $ 0.015439 \pm 0.000015 $  &  $ 0.11339 \pm 0.00011 $  &  $ 0.23738 \pm 0.00024 $  &  $ 0.7735 \pm 0.0010 $  &  $ 5.451 \pm^{0.021}_{0.019} $  \\
 &  $0.015439 \pm 0.000015$  &  $0.11339 \pm 0.00011$  &  $0.23735 \pm 0.00024$  &  $0.7733 \pm 0.0010$  &  $5.446 \pm 0.020$  \\ \hline
\multirow{2}{*}{$e$} &  $ 0.034 \pm^{0.022}_{0.021} $  &  $ 0.0023 \pm^{0.0025}_{0.0016} $  &  $ 0.073 \pm^{0.013}_{0.014} $  &  $ 0.046 \pm^{0.050}_{0.032} $  &  $ 0.0283 \pm^{0.0064}_{0.0066} $  \\
 &  $0.028 \pm^{0.022}_{0.019}$  &  $0.0023 \pm^{0.0025}_{0.0016}$  &  $0.072 \pm^{0.013}_{0.014}$  &  $0.080 \pm^{0.102}_{0.057}$  &  $0.0269 \pm^{0.0061}_{0.0065}$  \\ \hline
\multirow{2}{*}{$e_{avg}$} & $0.062 \pm^{0.192}_{0.039}$  &  $0.0194 \pm^{0.0048}_{0.0050}$  &  $0.0643 \pm^{0.0092}_{0.0107}$  &  $0.046 \pm^{0.050}_{0.032}$  &  $0.0283 \pm^{0.0064}_{0.0065}$  \\
& $0.061 \pm^{0.196}_{0.040}$  &  $0.0194 \pm^{0.0049}_{0.0047}$  &  $0.0638 \pm^{0.0094}_{0.0103}$  &  $0.080 \pm^{0.102}_{0.057}$  &  $0.0269 \pm^{0.0061}_{0.0065}$  \\ \hline
\multirow{2}{*}{$e\cos{\omega}$} &  $ 0.003 \pm^{0.024}_{0.023} $  &  $ -0.0000 \pm^{0.0018}_{0.0019} $  &  $ 0.069 \pm 0.013$  &  $ -0.020 \pm^{0.030}_{0.063} $  &  $ 0.0273 \pm^{0.0064}_{0.0066} $  \\
 &  $-0.003 \pm^{0.018}_{0.024}$  &  $-0.0000 \pm0.0018$  &  $0.068 \pm 0.013$  &  $-0.066 \pm^{0.066}_{0.108}$  &  $0.0258 \pm^{0.0062}_{0.0066}$  \\ \hline
\multirow{2}{*}{$e\sin{\omega}$} &  $ 0.023 \pm^{0.021}_{0.020} $  &  $ 0.0009 \pm^{0.0026}_{0.0014} $  &  $ 0.017 \pm 0.016 $  &  $ -0.006 \pm^{0.029}_{0.035} $  &  $ 0.0000 \pm 0.0072 $  \\
 &  $0.018 \pm^{0.021}_{0.018}$  &  $0.0008 \pm^{0.0028}_{0.0015}$  &  $0.018 \pm^{0.016}_{0.017}$  &  $0.002 \pm^{0.051}_{0.034}$  &  $0.0036 \pm 0.0071$  \\ \hline
\multirow{2}{*}{$\omega+M$ [degrees]} &  $ 111.56 \pm^{6.90}_{5.03} $  &  $ 327.23 \pm^{0.76}_{0.79} $  &  $ 361.49 \pm^{5.64}_{5.70} $  &  $ 328.00 \pm^{12.66}_{12.63} $  &  $ 176.69 \pm^{2.82}_{2.49} $  \\
 &  $112.38 \pm^{8.59}_{5.05}$  &  $327.18 \pm^{0.75}_{0.81}$  &  $358.19 \pm^{6.14}_{6.12}$  &  $324.76 \pm^{12.48}_{12.70}$  &  $176.32 \pm^{2.62}_{2.60}$  \\ \hline
 \multirow{2}{*}{$i$ [degrees]} &  $ 90.58 \pm^{3.57}_{4.35} $  &  $ 88.80 \pm^{25.46}_{23.89} $  &  $-$ & $-$  & $-$   \\
 &  $90.36 \pm^{3.96}_{4.66}$  &  $89.73 \pm^{24.49}_{24.54}$  &  &  &   \\ \hline
\multirow{2}{*}{$\Omega$ [degrees]} &  $ 353.88 \pm^{21.35}_{36.65} $  &  $ 0.00 \pm 0.00 $ & $-$  & $-$  & $-$ \\
 &  $352.44 \pm^{21.08}_{122.18}$  &  $0.00 \pm 0.00$  &  &  &  \\
\hline
\hline
\end{tabular}

\medskip
Time-averaged periods ($P_{avg}$) and eccentricities ($e_{avg}$) are also shown. The other parameters are based on our chosen RV epoch (the first Lick observation). The top value in each cell comes from Case 1 and the bottom value from Case 2.
\end{table*}

\begin{table*}
\caption{Estimates for the RV zero-point offsets and jitters described in \S \ref{sec:modelOfObs}. \label{tbl-4} }
\begin{tabular}{ccccc}
\hline
Parameter [\ms] & Lick & Keck & HET & HJST \\
\hline
\hline
\multirow{2}{*}{$C_{1,X}$} &  $1.09 \pm^{4.77}_{4.97}$  &  $-32.53 \pm^{0.91}_{0.89}$  &  $1.03 \pm^{0.68}_{0.71}$  &  $-0.24 \pm^{0.54}_{0.51}$  \\
& $0.09 \pm^{5.23}_{5.30}$  & $-32.36 \pm^{0.89}_{0.83}$  & $0.92 \pm^{0.75}_{0.72}$  & $0.23 \pm^{0.55}_{0.51}$  \\ \hline
\multirow{2}{*}{$C_{2,X}$} &  $21.34 \pm^{3.48}_{3.53}$  &  $-32.91 \pm 0.35 $  &  - &  - \\
& $21.56 \pm^{3.57}_{3.46}$  & $-33.13 \pm^{0.37}_{0.38}$  & - & - \\ \hline
\multirow{2}{*}{$C_{3,X}$} &  $-5.82 \pm^{0.84}_{0.83}$  &  - &  - &  - \\
& $-6.20 \pm^{0.82}_{0.79}$  & - & - & - \\ \hline
\multirow{2}{*}{$C_{4,X}$} &  $-3.17 \pm^{0.94}_{0.87}$  &  - &  - &  - \\
& $-2.86 \pm^{0.96}_{0.97}$  & - & - & - \\ \hline
\multirow{2}{*}{$C_{5,X}$} &  $-6.22 \pm^{0.43}_{0.42}$  &  - &  - &  - \\
& $-6.49 \pm 0.42$  & - & - & - \\ \hline
\multirow{2}{*}{$\sigma_{jit,X1}$} &  $2.37 \pm^{3.38}_{1.85}$  &  $3.46 \pm^{0.14}_{0.13}$  &  $4.66 \pm^{0.44}_{0.42}$  &  $4.85 \pm^{0.33}_{0.32}$  \\
& $3.01 \pm^{3.98}_{2.37}$  & $3.18 \pm^{0.15}_{0.14}$  & $3.85 \pm^{0.49}_{0.46}$  & $4.66 \pm 0.38$  \\ \hline
\multirow{2}{*}{$\sigma_{jit,X2}$} &  $6.06 \pm^{0.44}_{0.40}$  &  - &  - &  - \\
& $5.39 \pm^{0.48}_{0.47}$  & - & - & - \\ \hline
\multirow{2}{*}{$\sigma_{jit,X3}$} &  $6.75 \pm^{0.29}_{0.26}$  &  - &  - &  - \\
& $6.35 \pm^{0.32}_{0.31}$  & - & - & - \\

\hline
\hline
\end{tabular}

\medskip
The top value in each cell comes from Case 1 and the bottom value from Case 2.
\end{table*}

Lastly, we considered possible correlations amongst multiple RV observations.
In our standard \RUNDMC, the model of observations assumes the RV measurement errors are normally distributed and uncorrelated with one another.
This is usually an excellent approximation when observations are separated by one or more days.
For bright stars like 55 Cancri, the required exposure time is often less than a minute, which is not long enough to average over Solar-like p-mode oscillations.
In the past when multiple RV observations were taken sequentially, observers binned these observations and reported a single measurement in an attempt to average over the stellar noise and improve the RV precision.
The resulting coarse time series makes it difficult to probe orbital periods in the sub-day regime.
However, 55 Cancri A is a very bright RV target (V$\sim$6), and observations can be binned more frequently (every $\sim$10 minutes) with enough dedicated telescope time.
Unfortunately, this introduces the potential for significant correlations amongst back-to-back observations.
Rather than trying to model these correlations in detail, we consider two extreme cases: 1) the typical assumption that all RV errors are uncorrelated (hereafter Case 1) and 2) the measurement errors of observations taken within a maximum of 10 minutes from each other are perfectly correlated (hereafter Case 2).
To simulate the latter, we scale the uncertainties of back-to-back RVs by the square root of the number of observations in that set.
For example, the RV uncertainties in each of a set of 12 short cadence observations are scaled up by $\sqrt{12}=3.464$.
In reality, the true correlation amongst RV observations is somewhere in between these two extreme assumptions.
As we will show, the differences in the values for the majority of model parameters were negligible.
Table \ref{tbl-2} provides a convenient reference for the definitions of Case 1 and Case 2, as these terms will be used throughout the rest of the paper.
We perform two sets of \RUNDMC jobs based on Case 1 and Case 2.


\section{Results}
\label{sec:results}
We apply the \RUNDMC algorithm to 1,418 RV observations spanning an observing baseline of $\sim$23 years.
We obtain $\chi^2=1419\pm52$ and $\chi^2=1333\pm49$ for Cases 1 and 2 respectively.
Table \ref{tbl-3} lists our estimations of all the planetary parameters for both cases.
Our estimates for the zero-point offsets and jitters are shown in Table \ref{tbl-4}.

\begin{figure}
\centering
\includegraphics[width=0.9\columnwidth]{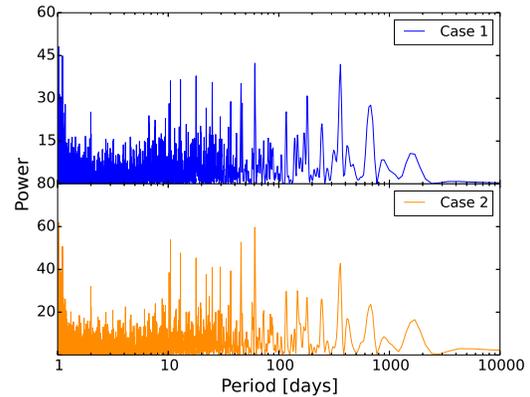}
\caption{ The periodogram of RV residuals for Cases 1 (top, blue) and 2 (bottom, orange). We do not probe the sub-day regime as we do not expect to find planets in addition to \E in this period domain. }
\label{fig:periodogram}
\end{figure}

We estimate the RV residuals by averaging the orbital elements in our RUN DMC ensemble for the last Markov chain generation and subtract the RV curve predicted by the n-body mode from our RV data.
The effective residual uncertainties are based on adding the RMS RV residuals and measurement uncertainties in quadrature.
In Figure \ref{fig:periodogram}, we plot the residual periodograms for Cases 1 and 2.
We find no obviously significant signals.
There are significant peaks in regions associated with common time sampling aliases, e.g. near one-day and one-year, but a peak near 60.7589 days stands out in both cases.
It is not immediately obvious whether a planet at this orbital period would be prone to instabilities due to the proximity of planets \B, \C, and \F.
We use a Keplerian MCMC \citep{Ford06} to fit the residual RVs with a one-planet model at the 60.7589 day peak.
The posterior distribution is multi-modal, the median eccentricity is large ($\sim$0.5), and the half-amplitude is $\sim$2\ms, less than any of our \jitter estimates.
We note that \citet{Baluev13} presented a highly parallelized Fourier decomposition algorithm that analyzed the \citet{Fischer08} velocities and uncovered a 9.8 day signal.
We do not detect such a signal in our extended dataset.

The following subsections describe our results for each of the known 55 Cancri planets and the orbital dynamics of the system.
Our results are based on posterior samples from the Markov chain output of \RUNDMC and these are fed into \MERCURY as sets of initial conditions for the dynamical simulations.


\subsection{The Transiting Planet, 55 Cancri e}
\label{sec:planete}
The high cadence of the RV time series provides excellent sampling of orbital periods, even those less than one day like planet \E.
We find the biggest difference between Case 1 and Case 2 is seen in planet \E's velocity amplitude, $K_e$ ($ 6.04 \pm 0.19$ and $6.12 \pm 0.20$ \ms) and therefore $m_e$.
When assuming Case 2, the median values of $K_e$ and $m_e$ shift to greater values relative to Case 1, compensating for the reduced RV information.
For $e_e$, both cases are consistent with zero.
Given the age of the star and the typical timescales for tidal circularization, one would expect $e_e\sim0$, but the presence of the other planets may maintain an excited orbit \citep{DawsonFabrycky10}.

We perform transit timing variation (TTV) simulations to address how significantly the outer four planets affect the orbit of planet \E \citep{Veras11,FordHolman07}.
The TTV amplitude, i.e. the maximum about of time which the orbit of \E deviated from a linear ephemeris, was on the order of seconds, much smaller than the observing cadence for binned observations with Spitzer or MOST.
Therefore, we expect its transit times to be very nearly periodic.

\begin{figure*}
\includegraphics[width=0.8\textwidth]{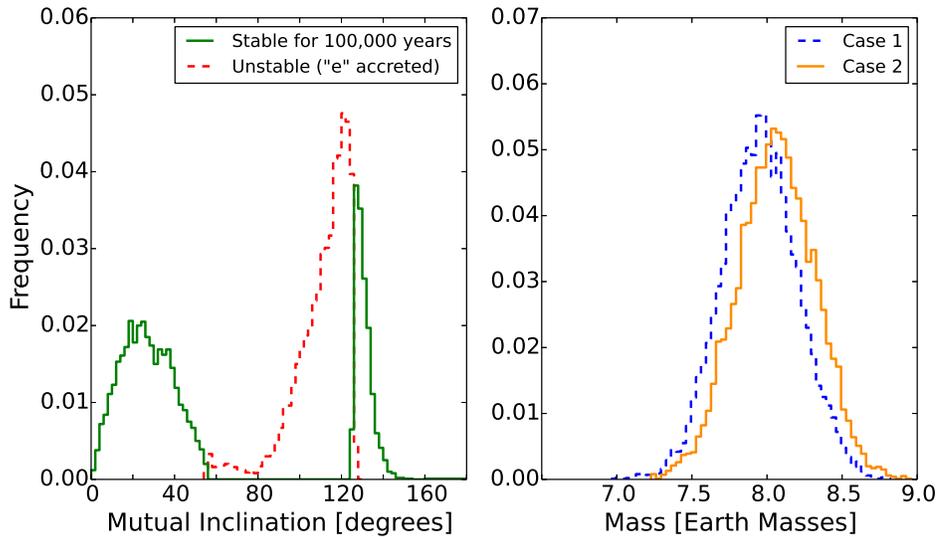}
\caption{ Properties of the inner-most planet, \E. \textbf{Left}: Two distributions of the mutual inclination between planet \E and the outer four (coplanar) planets for Case 2. In 10,000 dynamical simulations, 5,056 systems had the inner-most planet spiral into 55 Cancri A within $10^5$ years (normalized distribution in red/dashed). The remaining systems (normalized distribution in green/solid) showed no signs of instability, defined in \S \ref{sec:planetf}. \textbf{Right}: Distribution of planet mass for Case 1 (blue, dashed) and Case 2 (orange, solid). Mass estimates with respect to previous results are listed as follows: $8.63\pm0.35$  \citep{Winn11}, $8.37\pm0.38$ \citep{Endl12}, $7.99\pm0.25$ (Case 1), $8.09\pm0.26$ \Mearth (Case 2). }
\label{fig:planete}
\end{figure*}

Because we allowed planet \E to have a different $i$ and $\Omega$ than the outer planets, we can attempt to constrain the mutual inclination, \mutInc, with the rest of the planets.
Before we impose the constraint of dynamical instability, more than half of our models had planet \E in a retrograde orbit.
Because \E was found to be dynamically decoupled from the rest of the planets, these models are indistinguishable on the observing baseline from that of a prograde orbit.
By integrating these systems for 10$^5$ years, we find that most models with large \mutInc are dynamically unstable (Figure \ref{fig:planete}, left).
The outcome of the instability is the inner-most planet being accreted by the host star.
The time until accretion is rapid ($<$$10^5$ years) for all mutual inclinations that result in instability.
The instability is typically even more rapid ($<$$10^3$ years), with the exception of $\mutInc\sim125\degree$.
Note that we find no indication of instability for orbits with $\mutInc>125^o$.
We do not find any strong correlations between planet \E's stability and any other parameters.
Therefore, we obtain a new set of 10,000 posterior samples that do not include systems with 60$\degree < \mutInc < 125\degree$.
In Figure \ref{fig:planete}, there are overlapping stable and unstable solutions near 125$\degree$.
These solutions are unstable on longer timescales (to be discussed in \S \ref{sec:stability}), so we consider 125$\degree$ to be a conservative cutoff.
It is worth noting that when we include the effects of relativistic precession on planet \E, most of these retrograde orbits become unstable.
We find the timescale for this relativistic precession is comparable to that of the secular interactions.

In the right panel of Figure \ref{fig:planete}, we show distributions of planet mass for Cases 1 and 2.
The additional RVs plus the self-consistent dynamical model of the system provide an improved constraint on the mass and density of planet \E.
The median value is roughly 2-sigma lower than the \citet{Winn11} value and 1-sigma lower than the \citet{Endl12} value.
The uncertainty in mass is nearly $\sim$3\% in both cases.
Using the \citet{Winn11} planet-to-star radius ratio, we constrain the density of \E to $5.44\pm^{1.29}_{0.98}$ and $5.51\pm^{1.32}_{1.00}$ \gcm for Cases 1 and 2, respectively.
 
We do not include the effects of tidal dissipation on the 55 Cancri system, as the tidal dissipation rate is highly uncertain.
\citet{Bolmont13} find tidal dissipation timescales for $>$$10^5$ years, even for very high tidal efficiencies.
Therefore, tides will have a negligible effect on the system over the timescale of observations being analyzed.
They find tides affect the orbital evolution on timescales of $\sim$$3\times10^4$ years.
For most mutual inclinations that result in instability, the timescale to instability is significantly shorter than the predicted tidal evolution timescale.
Our measurement of the eccentricity for planet \E is inconsistent with significantly faster tidal dissipation.

Our orbital parameter estimates listed in Table \ref{tbl-3} represent a significant improvement on those of \citet{DawsonFabrycky10} and \citet{Endl12}.
We find the eccentricity of planet \E ($e_e=0.028\pm^{0.022}_{0.019}$) is significantly less than the estimate from \citet{DawsonFabrycky10} ($e_e=0.17$).
Thus, the current eccentricity places a much weaker constraint on the tidal efficiency than the previous analysis of \citet{Bolmont13} that assumed initial conditions from \citet{DawsonFabrycky10}.
\citet{Bolmont13} also found the initial conditions derived from \citet{Endl12} to be dynamically unstable, and thus not a realistic model.
Furthermore, since \citet{Endl12} assumed a circular orbit for planet \E, their orbital model is of limited value for exploring the tidal evolution of the system.
Nevertheless, many of the qualitative conclusions from \citet{Bolmont13} about the long-term eccentricity evolution of the system are likely applicable, since their measurements of the eccentricities for planets \B and \C are similar to those of our more detailed modeling.
Using the \citet{Endl12} initial conditions, \citet{Bolmont13} predict $e_e$ approaches an equilibrium where the eccentricity periodically varies from a nearly circular configuration to maximum eccentricity of $\sim$$5\times10^{-4}$ to 0.013.

\begin{figure}
\begin{center}
\includegraphics[width=1.0\columnwidth]{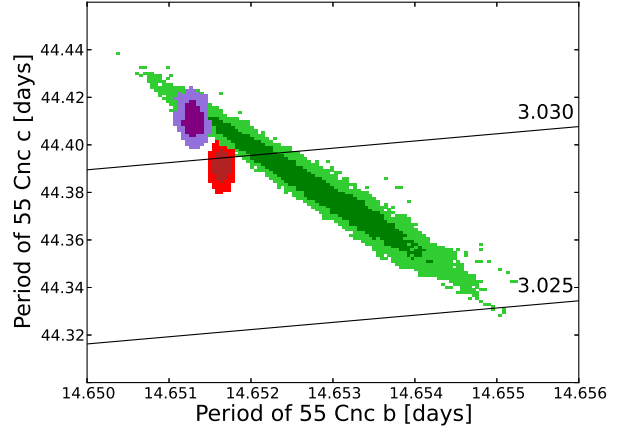}
\caption{Estimates of the period ratio of the \B and \C pair for Case 2. Two lines of constant period ratio (3.025, 3.030) are shown for reference. We draw 10,000 independent samples from a Keplerian MCMC (red). Choosing the RV epoch to be the first Lick observation, we draw 10,000 samples that are also vetted for stability (see Figure \ref{fig:planete}) from a \RUNDMC (green) analysis. For reasons stated in the text (\S \ref{sec:planetbc}), we compute time-averaged orbital period estimates of the raw \RUNDMC solutions over the $10^5$ year integration (purple). 1-$\sigma$ and 2-$\sigma$ credible contours are constructed, shown as the darker and lighter color respectively. }
\label{fig:periodRatio}
\end{center}
\end{figure}


\subsection{The Near Resonant Pair, 55 Cancri b and c}
\label{sec:planetbc}
In Figure \ref{fig:periodRatio}, we show 68.3\% and 95.4\% credible interval contours in $P_b$ and $P_c$ space for Case 2.
Most previous analyses for 55 Cancri assumed independent Keplerian orbits, but Figure \ref{fig:periodRatio} demonstrates the importance of n-body effects during the $\sim$23 years of observations.
Our n-body analysis adopts a Jacobi coordinate parameterization based on an epoch corresponding to the first Lick observation in February 1989.
Using \MERCURY, we extend the integration for $10^5$ years on a 10,000 dynamically ``stable'' models described in \S \ref{sec:planete} and find the maximum variability in the semi-major axes of planets \B and \C is comparable to or even exceeds the 68.3\% credible intervals of the respective semi-major axis distributions.
The epoch for initial conditions happened to be around this maximum, so our time averaged estimates for $P_b$ and $P_c$ are not centered on the contour of the raw \RUNDMC orbital periods.
The median values of the Keplerian and n-body solutions are separated by $\sim$6 $\times$ the standard deviation of the joint Keplerian distribution in $P_b$-$P_c$ space.

The three resonant angles associated with the 3:1 MMR were circulating for nearly all of our models, so these two planets are not in a mean-motion resonance.
This is consistent with the findings of \citet{Fischer08}, and from our large sample size, we can conclude that this result robust.

\begin{figure*}
\begin{center}
\includegraphics[width=0.8\textwidth]{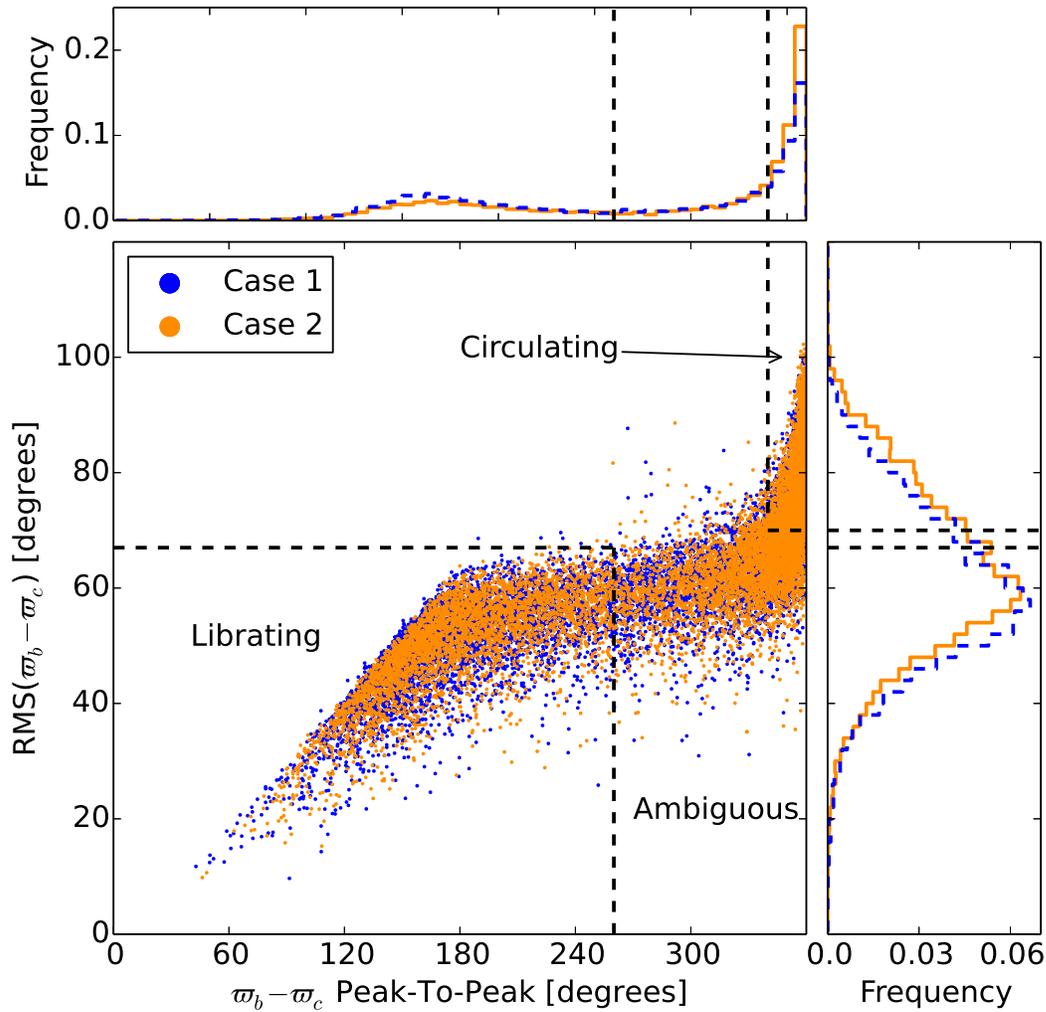}
\caption{ The secular behavior of 10,000 model systems for Case 1 (blue, dashed) and Case 2 (orange, solid) as discussed in \S \ref{sec:planetbc}. The horizontal axis shows the peak-to-peak variation in the secular angle ($\varpi_b-\varpi_c$) only when the eccentricity of either planet was greater than 0.001. The vertical axis shows the RMS of the secular angle variation. Librating systems typically had RMS $<$$67\degree$ and peak-to-peak variation $<$$260\degree$. Circulating systems typically had RMS $>$$70\degree$ and peak-to-peak variation $>$$340\degree$. Those that did not fall into either of these categories displayed an array of secular behavior, ranging from large amplitude libration with short term variations to long-term ``nodding'' \citep{Ketchum13}. }
\label{fig:libration}
\end{center}
\end{figure*}

Nevertheless, they are near enough to the 3:1 MMR that they can interact significantly and exhibit interesting interactions.
In particular, they might develop a secular lock, a configuration where both periastrons precess at the same time-averaged rate, and the gravitational kicks from frequent conjunctions cause libration of the angle between their pericenter directions.
We consider a subset of 10,000 models from Case 1 and Case 2 and track the orbital evolution for $10^5$ years using \MERCURY.
We compute the root-mean-square (RMS) of the angle $\varpi_b-\varpi_c$, using six different phase domains for the angle (0 to $2\pi$, $-5\pi/3$ to $1\pi/3$, $-4\pi/3$ to $2\pi/3$, $-\pi$ to $\pi$, $-2\pi/3$ to $4\pi/3$, $-\pi/3$ to $5\pi/3$).
We find the smallest RMS value and report the value centered on the domain with the smallest RMS value.
The vast majority of cases were centered about $\pi$ (180$\degree$).
While many of these are librating, some systems that are circulating but evolve more slowly when the secular angle is near $\pi$ might also be classified as librating.
Additionally, the short-term interactions with the other planets can cause the long-term librating behavior of \B and \C to occasionally circulate.

We considered multiple methods for estimating the libration amplitude for each of these models.
The naive approach would be to simply compute half of the peak-to-peak variation.
However, the secular interactions between planets \B and \C sometimes pushed them toward very low eccentricity. 
For nearly circular orbits, short-term perturbations can cause $\varpi_b-\varpi_c$ to take on extreme values even for systems undergoing secular libration.
Thus, the peak-to-peak method overestimates the libration amplitude for many systems.
This was remedied by applying an eccentricity cutoff to filter out values of $\varpi_b-\varpi_c$ when $e_b$ or $e_c<0.001$.
We found the shape of the resulting libration amplitude distribution was strongly dependent on our chosen cutoff value.
In light of this, we tried other metrics more robust to outliers, such as the RMS and median absolute deviation (MAD).
For a system undergoing small amplitude, sinusoidal libration, RMS$\times \sqrt{2}$ and MAD$\times \sqrt{2}$ are excellent approximations for the libration amplitude.

We found a wide range in $\varpi_b-\varpi_c$ behavior, so there was not one metric that could accurately differentiate libration from circulation.
Instead, we compute the RMS and peak-to-peak variation of the secular angle including only values where both eccentricities exceed 0.001.
Figure \ref{fig:libration} shows a scatter plot of the RMS and peak-to-peak variation for 10,000 ``stable'' models from \S \ref{sec:planete} for each case.
Based on visual inspection, we find that small amplitude librations typically have RMS($\varpi_b-\varpi_c$) $<$$67\degree$ (corresponding to a libration amplitude less than 90\degree) and a peak-to-peak variation $<$$260\degree$.
The median libration amplitude of this subset of models is $51\degree\pm^{6\degree}_{10\degree}$ (68.3\% credible interval) for both Cases 1 and 2.
Circulating systems typically have RMS $>$$70.0\degree$ and peak-to-peak variation $>$$340\degree$, but many of these were ``nodding'', meaning that $\varpi_b-\varpi_c$ exchanges between cycles of libration and circulation \citep{Ketchum13}.
Those that do not meet either of these aforementioned criteria fall into the ``ambiguous'' category.
By inspection, many of the ambiguous systems have large libration amplitudes with short-term perturbations, while others show long-term nodding.
Under our empirically derived definitions for libration and circulation, we find the fraction of stable systems undergoing the following types of secular behavior: librating, 44.6\% (36.5\%); circulating/nodding, 20.3\% (28.2\%); and ambiguous, 35.1\% (35.3\%) for Case 1 (Case 2).

Case 2 favors a larger eccentricity value for planet \F than Case 1.
This could suggest that the closer periastron passages of \F in Case 2 disrupts the secular lock of the near resonant pair, causing a greater fraction of modeled systems to circulate.


\subsection{The Habitable Zone Planet, 55 Cancri f} 
\label{sec:planetf}
When the three planet model of 55 Cancri was announced by \citet{Marcy02}, there was a noticeably wide semi-major axis gap between planets \C and \D.
An unseen planet could reside in this dynamically stable gap, possibly one massive enough to induce a clear RV signal \citep{Raymond08}.
\citet{Fischer08} announced the fifth planet, \F, a sub-Jovian mass planet on a Venus-like orbit, residing in the classical habitable zone.
Because it induces the weakest RV signal of all the 55 Cancri planets, its eccentricity is not well constrained; in Case 1, its eccentricity is consistent with circular, but in Case 2, it has a substantial range in eccentricity with a median of $\sim$0.1.
In the latter case, the stellar flux it receives over time could vary substantially.


\begin{figure}
\begin{center}
\includegraphics[width=1.0\columnwidth]{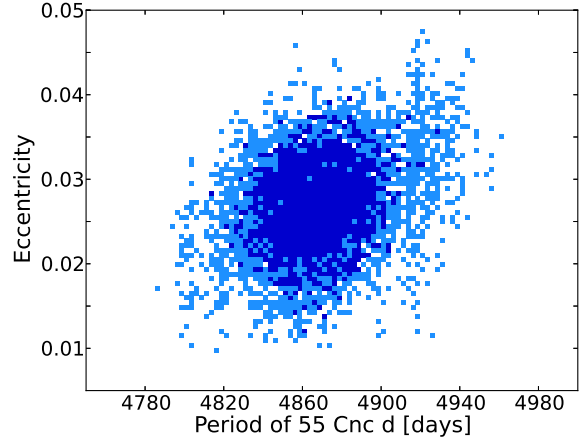}
\caption{Marginal posterior density for time-averaged period and time-averaged eccentricity of 55 Cancri \D, based on Case 2. }
\label{fig:planetd}
\end{center}
\end{figure}

\subsection{The Jupiter Analog, 55 Cancri d}
\label{sec:planetd}
In terms of its measured period and eccentricity, planet \D is the closest Jupiter analog to date.
The \citet{Endl12} estimates are $P_d=4909\pm30$ days and $e_d=0.02\pm0.008$, compared to that of Jupiter: $P_J\approx4333$ days and $e_J\approx0.049$.
Our posterior samples not only improve the parameter uncertainties, but also nudge the estimated period and eccentricity closer to that of Jupiter ($P_d\approx4867$ days and $e_d\approx0.0269$).
Although \D has yet to undergo two full period cycles, we are able to infer small but measurable eccentricity (Figure \ref{fig:planetd}).


\subsection{Long-Term Stability}
\label{sec:stability}
We tested the long-term stability of the system, excluding the wide binary companion.
We modified \MERCURY such that if any planet's semi-major axis changes by more than 50\% of its original value ($\left|[a_{final}-a_{initial}]/a_{initial}\right|>0.5$), then the simulation stops and is tagged as being unstable.
First, we evolved 1,000 five-planet models for $10^6$ years, and identified 9.4\% as unstable within $60\degree < \mutInc < 125\degree$.
The stability was most sensitive to the mutual inclination between the orbital plane of planet \E and the common orbital plane of the remaining planets on this longer timescale.
Over 90\% of the unstable solutions were those where \mutInc $\sim$ 125\degree, and planet \E was accreted by its host star.
However, we still found stable solutions with planet \E orbiting retrograde.
Thus, dynamical stability provides constraints on \mutInc, as described in \S\ref{sec:planete}.  
For systems with \mutInc $<$$60\degree$, the inner planet was dynamically decoupled from the outer four planets.

Next, we evolved 1,000 four-planet models (excluding the inner-most planet) for $10^8$ years.
None of the 1,000 model systems tested were identified as unstable.
Thus, the extensive RV observations now provide such tight constraints on the planet masses and orbital parameters that essentially all allowed nearly coplanar solutions are dynamically stable.  

Prior to our final analysis, we preformed several preliminary analyses based on a different set of observations (fewer RV measurements, Lick velocities from \citet{Fischer08} rather than improved reduction in 2014).
The posterior samples included some models with the eccentricity of planet \F exceeding 0.3.
For many of these sets of initial conditions, systems would become unstable, recognized by a significant change in semi-major axis of planet \C or \F on a timescale of $\sim$$10^6-10^8$ years. 
We conclude that the tighter constraint on the eccentricity of planet \F based on our final analysis was responsible for ensuring that none of the 1,000 model systems from our final analysis showed signs of dynamical instability.


\section{Discussion}
\label{sec:discussion}
We report the first self-consistent Bayesian analysis of the 55 Cancri system that uses n-body integrations to account for the planets' mutual gravitational interactions.
By combining a rigorous statistical analysis, dynamical model and improved observational constraints, we obtain the first set of five-planet models that are dynamically stable.  
We considered two extreme cases where the high cadence, unbinned RV measurements were treated as independent (Case 1) or perfectly correlated (Case 2).
In both cases, the RV residuals show no immediately convincing signals due to an additional sixth planet.

Informative priors based on precise photometry from MOST \citep{Winn11} drastically narrow the possible orbital inclination range of planet \E.
The planet-planet interactions amongst the remaining planets provide a loose inclination constraint just based on the RVs.
Combining all this information yields a mutual inclination estimate.
Under the assumption of relatively short-term dynamical stability ($10^5$ years), we find planet \E cannot be highly misaligned with the outer four planets.
However, there are some stable configurations where \E orbits retrograde as long as the mutual inclination is nearly retrograde ($125<i\leq180\degree$).
To be consistent with our photometric knowledge, we utilize the MOST planet-to-star radius ratio measurement to obtain a density estimate of $5.44\pm^{1.29}_{0.98}$ and $5.51\pm^{1.32}_{1.00}$ \gcm for Cases 1 and 2 respectively.
Assuming the derived mass for \E is insensitive to our priors on the central transit time, transit duration, and ingress duration, the radius ratio estimate from \citet{Gillon12} gives alternative densities of $4.28\pm^{0.65}_{0.55}$ and $4.34\pm^{0.66}_{0.55}$ \gcm for Cases 1 and 2 respectively.

We find that planets \B and \C are not in the 3:1 MMR, as all of the associated resonant angles are circulating.
We find that between a 36-45\% chance for the pericenter directions of planets \B and \C to exhibit secular lock with $\varpi_b-\varpi_c$ librating about 180$\degree$.
The behavior of this angle is only weakly constrained by the present observations, with a substantial fraction of solutions resulting in circulating or nodding of the secular angle.

Using the latest RV datasets, we find the vast majority of, if not all, models of the outer-four planets are long-term stable.
We expect that tidal effects and gravitational perturbations due to planet \E are negligible over these timescales.

Despite the large dimensionality, short integration timestep, and a lengthy mixing time, we have obtained a set of effectively independent posterior samples available to the exoplanet community for more detailed future studies.
The improved mass and density estimates for \E will certainly provide new insight to the interior bulk composition.
If it turns out that planet \B has an extended atmosphere that grazes its host star \citep{Ehrenreich12}, our updated orbital period and mass estimates will be valuable in modeling the planet's atmosphere and potentially providing a sharper orbital inclination estimate.
Astrometric observations of 55 Cancri A with the Hubble Space Telescope traced out a small arc, presumably dominated by orbit of planet \D \citep{McGrath03}.
In combination with RVs, the orbital inclination of the outer-most planet was estimated to be $53\pm6.8\degree$ \citep{McArthur04}, but this relies on an accurate assessment of \D's orbital period, which has changed significantly since then.
We recommend a re-analysis of the HST data with the new orbital model and incorporating the M dwarf companion to improve the astrometric constrain on its orbital inclination.

For the advancement of the exoplanet field in general, we are releasing the posterior samples from this analysis and encourage others to do the same for future announcements or updates to individual planetary systems.
Posterior samples for Case 1 and Case 2 are available as Supporting Information with the online version of the article.
In particular, it has now become practical to use hierarchical Bayesian models to infer properties of the exoplanet population.
Understanding the nature behind the uncertainties in orbital period, mass, eccentricity, etc. beyond the 1-$\sigma$ values is crucial for such a study.

\section*{Acknowledgements}
We would like to thank our referee Sean Raymond for his helpful comments on the manuscript.
B.E.N. thanks Eric Feigelson and Brad Hansen for insightful discussions of that helped strengthen the paper.
We thank Geoff Marcy and the entire of the California Planet Survey team for their long-term commitment to high-precision RVs for the 55 Cancri system.
We also thank Stan Dermott for his contribution to this project.
This research was supported by NASA Origins of Solar Systems grant NNX09AB35G and NASA Applied Information Systems Research Program grant NNX09AM41G.
The authors acknowledge the University of Florida High Performance Computing Center for providing computational resources and support that have contributed to the results reported within this paper.
The Center for Exoplanets and Habitable Worlds is supported by the Pennsylvania State University, the Eberly College of Science, and the Pennsylvania Space Grant Consortium.
We extend special thanks to those of Hawai'ian ancestry on whose sacred mountain of Mauna Kea we are privileged to be guests.  
Without their generous hospitality, the Keck observations presented herein would not have been possible.

\bibliographystyle{aa} 
\bibliography{references}

\clearpage

\label{lastpage}
\end{document}